\documentstyle[epsf,emulateapj]{article}

\begin{document}

\title{The North Ecliptic Pole Supercluster}
\author{C.R.\ Mullis\altaffilmark{1}, J.P.\ Henry\altaffilmark{1},
        I.M.\ Gioia\altaffilmark{1,2},\\ H.\ B\"{o}hringer\altaffilmark{3},
        U.G.\ Briel\altaffilmark{3}, W. Voges\altaffilmark{3},
        and J.P.\ Huchra\altaffilmark{4}}
\altaffiltext{1}{Institute for Astronomy, University of Hawai`i, 
		 2680 Woodlawn Drive, Honolulu, HI 96822, USA, mullis@ifa.hawaii.edu}
\altaffiltext{2}{Istituto di Radioastronomia CNR, via Gobetti 101, 
                 Bologna, I-40129 ITALY}
\altaffiltext{3}{Max-Planck Institut f\"u{r} extraterrestrische Physik, 
                 Giessenbachstrasse 1603, Garching, D-85741, GERMANY}
\altaffiltext{4}{Harvard-Smithsonian Center for Astrophysics, 60 Garden Street, Cambridge, MA 02138, USA}

%\slugcomment{submitted to ApJ}

\begin{abstract}

We have used the ROSAT All-Sky Survey to detect a known supercluster
at $z=0.087$ in the North Ecliptic Pole region.  The X-ray data
greatly improve our understanding of this supercluster's
characteristics, approximately doubling our knowledge of the
structure's spatial extent and tripling the cluster/group membership
compared to the optical discovery data.  The supercluster is a rich
structure consisting of at least 21 galaxy clusters and groups, 12
AGN, 61 IRAS galaxies, and various other objects.  A majority of these
components were discovered with the X-ray data, but the supercluster
is also robustly detected in optical, IR, and UV wavebands.  Extending
$129 \times 102 \times 67~(h_{50}^{-1}\ $Mpc)$^{3}$, the North
Ecliptic Pole Supercluster has a flattened shape oriented nearly
edge-on to our line-of-sight.  Owing to the softness of the ROSAT
X-ray passband and the deep exposure over a large solid angle, we have
detected for the first time a significant population of X-ray emitting
galaxy groups in a supercluster.  These results demonstrate the
effectiveness of X-ray observations with contiguous coverage for
studying structure in the Universe.
 
\end{abstract}


\section{Introduction} 

Superclusters (SCs) are the ultimate manifestation of hierarchical
large-scale structure (LSS) in the Universe.  With a median major
axis of $\sim150~h_{50}^{-1}\ $Mpc (Jaaniste et al.\ 1998), they
are the most massive and the largest possibly bound structures known.
``Great walls'' of galaxies have been discovered that are larger than
what one normally calls a SC, but they are almost certainly not
gravitationally bound (Geller \& Huchra 1989).  Observations and
numerical simulations suggest that a rich hierarchy of constituents
trace a SC. Massive galaxy clusters are linked by smaller galaxy
groups in the high-density regions, while filamentary distributions of
galaxies fill out the expansive, low-density domains.  SCs are not
isolated entities but the top-level components of a ``cosmic web'' of
LSS (Tully 1987; Bond, Kofman, \& Pogosyan 1996; Einasto et al.\
1997).

Despite their ubiquity in our overall impression of the Universe, SCs
are nonetheless exceedingly rare relative to other types of objects.
The extreme properties of SCs are the results of the initial
conditions of the nascent universe developed by structure formation
mechanisms.  SCs are so large that the crossing time of the member
clusters is greater than the Hubble time.  Hence the imprint of the
initial conditions has not been erased by internal motions.  Therefore
understanding the nature of SCs is of great cosmological importance.

Much effort has been expended to locate SCs, primarily in the optical
regime.  Galaxy surveys provide nearly all-sky sampling of the local
universe out to $z \sim 0.03$ (e.g. CfA, Huchra, Vogeley, \& Geller
1999), while more restricted slice surveys (e.g. LCRS, Shectman et
al.\ 1996) reach to $z \sim 0.2$.  Galaxy clusters are more sparsely
distributed but efficiently probe SCs to intermediate redshifts ($z
\sim 0.3$) across essentially the entire sky (e.g.\ Bahcall \& Soneira
1984; Batuski \& Burns 1985; Einasto et al.\ 1997; Ebeling et al.\
1998).

Here we report on the successful application of X-ray data, featuring
high sensitivity over a large, contiguous solid angle, to characterize
SCs and LSS.  We have used the ROSAT All-Sky Survey to detect a
substantial SC in the North Ecliptic Pole (NEP) region -- a structure
discovered in the optical by Batuski \& Burns (1985) with subsequent
X-ray detections by Burg et al.\ (1992). We have for the first time
detected a significant number of X-ray emitting galaxy groups in a SC
yielding a greatly enhanced understanding of its extent and content.
These X-ray data combined with optical, IRAS, and HST/UV observations
make the NEP Supercluster (NEPSC) one of the most broadly detected
superstructures known.

Throughout this letter we assume a Hubble constant of $H_{0} = 50\
h_{50}~$km~s$^{\rm -1}~$Mpc$^{\rm -1}$ and a deceleration parameter of
$q_{0} = 0.5$, however the major conclusions do not change for other
reasonable cosmologies.  Co-moving distances are reported unless
otherwise indicated.

\section{Observations}

\subsection{ROSAT NEP Survey}

The ROSAT All-Sky Survey (RASS) is the first all-sky survey with an
imaging X-ray detector (Tr\"{u}mper 1983; Voges et al.\ 1999). The NEP
region of the RASS possesses the deepest exposure ($t_{\rm max} \sim
40 $ ks) and consequently the greatest sensitivity (median flux limit,
$f_{\rm x}$ = 7.8 $\times$ 10$^{\rm -14}$ erg s$^{\rm -1}$ cm$^{\rm -2}$
[0.5--2.0 keV]) of the entire survey (Voges et al.\ 2001).  We have
identified the physical nature of 99.6\% of the 445 X-ray sources with
fluxes measured at greater than $4\sigma$ significance in the
{9$^{\circ}$ $\times$ 9$^{\circ}$} survey region centered on the NEP
(Henry et al.\ 2001).  Of these, 64 (14.4\%) are galaxy clusters and
groups. X-ray luminous clusters are detected out to $z=0.81$, while
low-luminosity groups of galaxies are well-sampled to $z \approx
0.15$.

Our primary motivation for pursuing the ROSAT NEP Survey was to
construct a complete sample of X-ray selected galaxy clusters, and use
it to examine X-ray cluster luminosity evolution and to characterize
LSS.  The initial evolution results are presented by Gioia et al.\
(2001), while the first findings relevant to LSS are presented in this
Letter.  We emphasize the cluster sample is unique in that it is both
deep in flux sensitivity and contiguous on the sky.

In the top panel of Fig.\ 1 we plot the observed ROSAT NEP cluster
redshift distribution along with the expected distribution for a
spatially homogeneous, non-evolving cluster population.  The latter is
calculated by folding the local cluster X-ray luminosity function
(XLF, Ebeling et al.\ 1997; De Grandi et al.\ 1999; H.\ B\"{o}hringer
2000, private communication) through the ROSAT NEP Survey selection
function.  The observed redshift distribution shows a striking feature
at $z=0.087$.  Sixteen clusters lie in the redshift interval $0.07 \le
z \le 0.10$, 25\% of the entire cluster sample.  Depending on which
XLF determination is used, this interval is 3.3 to 4.3 times more
populated than expected which is significant at the $3.9\sigma$ to
$4.6\sigma$ level.

Active galactic nuclei (AGN) are the dominant class of X-ray emitters
in the ROSAT NEP survey (49.0\% of all sources) and they are another
means for examining LSS.  Though they are more sporadic markers than
clusters, they reach to greater depths ($z_{\rm max}\sim 4$).  A
$2.2\sigma$ density enhancement around $z=0.087$ is also seen in the
redshift distribution of ROSAT NEP AGN.  The middle panel of Fig.\ 1
shows the observed distribution compared to a spatially homogeneous,
evolving AGN population.  Twelve AGN are observed in the redshift
interval $0.07 \le z \le 0.10$ where 5.6 are predicted from the AGN
XLF (Miyaji, Hasinger, \& Schmidt 2000).  Note, an additional peak
($1.85\sigma$) in the AGN redshift distribution at $z=0.32$ is
comprised of two separate groups of AGN.

Using \em only \em the ROSAT X-ray data and supporting optical
observations from within the NEP survey boundaries, we have detected a
significant LSS consisting of \em at least \em 16 galaxy clusters and
groups, plus 12 AGN, one BL Lac, and an isolated elliptical galaxy
with an X-ray luminosity commensurate with that of a poor cluster.
Note there are additional SCs and clusters of AGN in the ROSAT NEP
data.  These structures along with spatial correlation analyses for
the cluster and AGN samples will be discussed in forthcoming papers.

\subsection{Other Relevant Optical/X-ray/IR/UV Observations}

The presence of LSS in the vicinity of the NEP was first noted in a
pioneering study by Batuski \& Burns (1985). They produced a finding
list of SC candidates (SCC) based on a percolation analysis of Abell
clusters, and found an association of six clusters (A2301, A2304,
A2308, A2311, A2312, A2315) approximately 5$^{\circ}$ from the NEP
(SCC \#47).  Only two of these clusters had spectroscopic redshifts at
the time, the remainder having photometric estimates.  More recent
percolation studies, working with improved redshift data, recovered
the SC candidate (SC\#34/2, Zucca et al.\ 1993; \#97, \#170, Einasto
et al.\ 1994, 1997), with the latter two analyses noting an additional
cluster A2295.  Four of the Abell clusters (A2295, A2301, A2304,
A2308) lie within the ROSAT NEP survey boundaries and are easily
detected in X-rays.  The other three (A2311, A2312, A2315), though
barely outside the NEP region, are detected in the ROSAT Bright Source
Catalog (BSC, Voges et al.\ 1999).

During the early phase of the ROSAT mission Burg et al.\ (1992)
combined a 50 ks pointed observation with all-sky scans to identify
five X-ray clusters and groups between redshifts 0.08 and 0.09 within
1.5$^{\circ}$ of the NEP. Except for one very low luminosity group
discovered in the pointed data (Hasinger, Schimdt, \& Tr\"{u}mper
1991) which is below our flux limit, our survey recovers these
detections.  We examined all X-ray sources in the ROSAT BSC within
10$^{\circ}$ of the known SC members.  There are no additional
candidate galaxy clusters or groups in the redshift interval $0.07 \le
z \le 0.10$. The nearest candidate (RX\,J1827.6+6135, $z = 0.1014$,
B\"{o}hringer et al.\ 2000) is just outside both the NEP survey
region and the nominal SC redshift boundary, and is not included in
our analysis.

The IRAS satellite mapped the infrared sky in a mode similar to ROSAT,
and hence also has enhanced sensitivity at the NEP.  Ashby et al.\
(1996) tested models of starburst galaxy evolution using a sample of
IRAS galaxies within 1.5$^{\circ}$ of the NEP and noted the strong
effect of LSS.  Fifteen galaxies, 20\% of their IR-selected sample,
lie between redshifts 0.085 and 0.090 with $\bar{z} = 0.088$.
Rinehart et al.\ (2000) have expanded the IRAS survey to 56
deg$^2$. This extended IRAS sample ({Fig.\ 1}, lower panel) has 61
galaxies in the redshift interval $0.07 \le z \le 0.10$ where 30.9 are
expected representing a $4.7\sigma$ excess. These IRAS sources are
predominantly spirals found outside the dense cluster environment. The
galaxies form a sheet estimated to be three times as dense as the
Great Wall (Geller \& Huchra 1989).

Additional evidence of structure in the NEP region is a Ly$\alpha$
absorber detected at $z_{\rm abs}=0.08910$ in the direction of a
well-known QSO H1821+643 ($z_{\rm em}=0.297$).  Tripp et al.\ (1998)
identified this absorbing system during a study of low-redshift weak
Ly$\alpha$ absorbers using HST UV spectroscopy.  This QSO was
originally discovered via its high X-ray luminosity and is the
brightest source in the ROSAT NEP Survey.

\section{Discussion}

A three-dimensional view of the conical NEP survey volume, showing the
distribution of clusters, groups, and AGN out to $z=0.3$, is presented
in Fig.\ 2 (right).  The NEPSC is clearly visible as an
over-density of X-ray emitters at $z=0.087$ ($D_{L} = 533
~h_{50}^{-1}\ $Mpc).  This complex is relatively disconnected, with a
candidate void lying on the high redshift side ($z \sim 0.12$).  The
nearest non-SC clusters are RX\,J1827.6+6135, A2293, and A2275 at
respectively 64.7, 87.2, and $121.8 ~h_{50}^{-1}\ $Mpc.

We use the non-weighted inertia tensor for the NEPSC groups and
clusters to determine the principal axes and estimate the size, shape,
and orientation of the structure.  The maximal extents in three
orthogonal directions are $129 \times 102 \times 67 ~(h_{50}^{-1}\
$Mpc)$^{3}$ with a 12$^{\circ}$ extent on the sky.  The major and
minor axes are oriented 42$^{\circ}$ and 85$^{\circ}$ relative to our
line-of-sight.  The NEPSC is basically an oblate ellipsoid seen nearly
edge-on.  Consequently, the effect of flattening the SC in the
line-of-sight direction in going from real space to redshift space is
less important than would be in a face-on scenario such as with the
Great Wall.

All eight Abell clusters associated with the NEPSC are detected in
X-rays with luminosities in the range of 0.2--3.6 $\times~10^{44}~
h_{50}^{-2}$\ erg~s$^{-1}$ [0.5--2.0 keV].  Note that three clusters
discovered by the ROSAT NEP Survey in this luminosity range are
missing from the Abell catalog, indicative of the incompleteness
inherent to optical selection.  Furthermore, there are an additional
10 groups of galaxies with X-ray luminosities in the range 2--9
$\times~10^{42}~ h_{50}^{-2}$\ erg~s$^{\rm -1}$ [0.5--2.0 keV].  This
is the first time a large sample of X-ray emitting galaxy groups has
been identified in a SC.

Fig.\ 2 (middle) shows a close-up view of the NEPSC where the
groups and clusters have been linked together by a minimum spanning
tree.  Minimum, median, and maximum link distances are respectively
5.6, 24.2, and $40.5~h_{50}^{-1}\ $Mpc.  The massive Abell clusters
that initially delineated the SC are linked together by newly
discovered, less massive X-ray clusters and groups which bridge the
region between the previously known SC and a very rich cluster of
galaxies (A2255) at the same distance, but not previously associated
with this structure. The new realization of the NEPSC has doubled in
physical size and tripled in cluster content. If this SC is
characteristic of others, then SCs can be substantially larger than
what is shown by the Abell clusters alone.

Filaments are the fundamental element of cosmic structure present in
all modern numerical simulations of LSS.  The NEPSC is oriented in
Fig.\ 2 (middle) to display a possible filament containing two groups
and two clusters.  The length of the filament, the distance between
A2304 and NEP 4990, is $20.7 ~h_{50}^{-1}\ $Mpc. We also show in Fig.\
2 (left) a filament from a Virgo consortium $\Lambda$+CDM ($z=0$) numerical
simulation (Jenkins et al.\ 1998) of LSS selected to resemble the NEP
structure.  The NEPSC filament is potentially the best match yet
observed in X-rays to what is seen in the simulations (see Briel \&
Henry 1995 and Scharf et al.\ 2000 for previous observations).

\section{Summary}

The fortunes of experimental design and serendipity have converged at
the NEP to reveal a prominent example of LSS in the Universe.  The
NEPSC ($z=0.087$) is one of the most robustly sampled superstructures
known --- 21 X-ray emitting clusters of galaxies demark the highest
density regions while 12 X-ray selected AGN and over 60 IRAS galaxies
trace out the lower density domains.  The $129 \times 102 \times
67~(h_{50}^{-1}\ $Mpc)$^{3}$ complex is strongly detected in X-ray,
optical, IR, and UV wavebands.  Detailed follow-up observations with
Chandra, XMM-Newton and ground-based facilities will permit us to examine
the dynamics of the structure, the evolutionary state of its
constituents, and the degree of alignments between the member
clusters.  The SC's edge-on orientation makes it an attractive
candidate in which to look for filaments of diffuse X-ray emitting
gas.

We have combined contiguous coverage with high sensitivity in X-rays
to efficiently detect objects in a SC ranging from AGN and poor galaxy
groups to rich galaxy clusters.  This is the first time a sizable
number of X-ray emitting galaxy groups has been identified in a SC,
thus yielding a better reckoning of a SC's girth and content.  The
ROSAT NEP Survey demonstrates the effectiveness of X-rays for studying
LSS and advocates a next generation, large format X-ray survey mission
to propel this field of study to the next level.

\acknowledgements 

It is a pleasure to thank Harald Ebeling, Brent Tully, and G\"{u}nther
Hasinger for fruitful discussions, and Karen Teramura for graphics
assistance. The kind support of the UH TAC and the expertise of the
Mauna Kea Observatory personnel are gratefully acknowledged.  We also
thank Stephen Rinehart for providing access to the extended IRAS NEP
sample prior to publication. C.R.M.  acknowledges partial financial
support from the NASA Graduate Student Researchers Program
(NGT5-50175) and the ARCS Foundation.  Support has also come from the
NSF (AST91-19216, AST95-00515), NASA (GO-5402.01-93A,
GO-05987.02-94A), the Smithsonian Astrophysical Observatory, NATO
(CRG91-0415), and the Italian ASI-CNR.

\newpage
\begin{figure*}
\vspace*{-5mm}  \epsfxsize=0.75\textwidth \hspace*{2cm} \epsffile{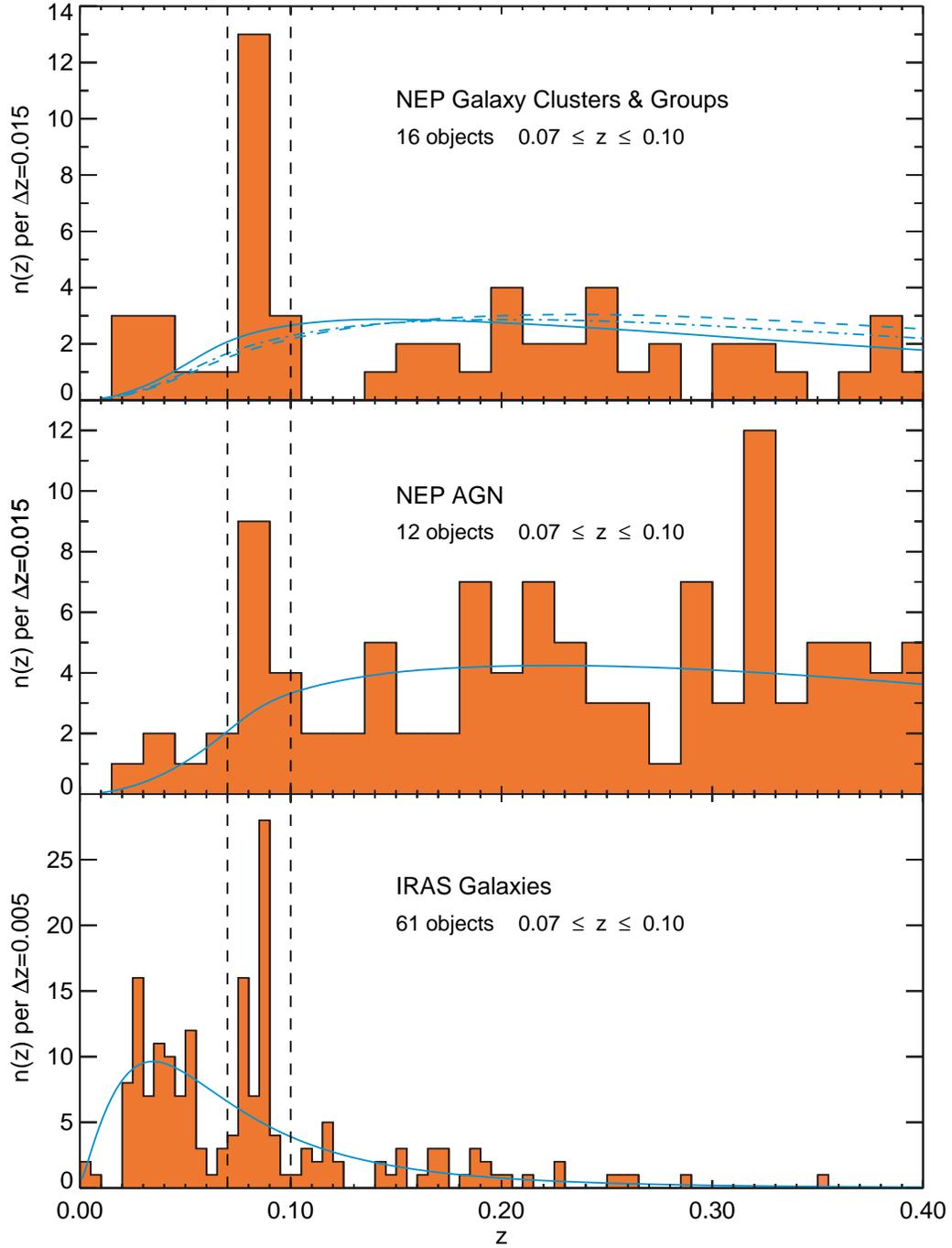}
\figcaption{Redshift distributions for ROSAT NEP clusters \& groups,
ROSAT NEP AGN, and IRAS galaxies.  The NEPSC redshift range is
indicated by vertical dashed lines at $z=0.07$ and $z=0.10$.  The
expected distributions of clusters are based on the BCS, RASS1BS, and
REFLEX XLFs (solid line - Ebeling et al.\ 1997; dashed line - De
Grandi et al.\ 1999; dot-dashed line - H.\ B\"{o}hringer 2000, private
communication).  The expected AGN distribution is based on the XLF of
Miyaji, Hasinger, \& Schmidt (2000).  The IRAS galaxy distribution, as
well as the expected distribution, are from Rinehart et al.\ (2001).}
\end{figure*}

\newpage
\begin{figure*}
\vspace*{-1mm} \epsfxsize=1.0\textwidth \epsffile{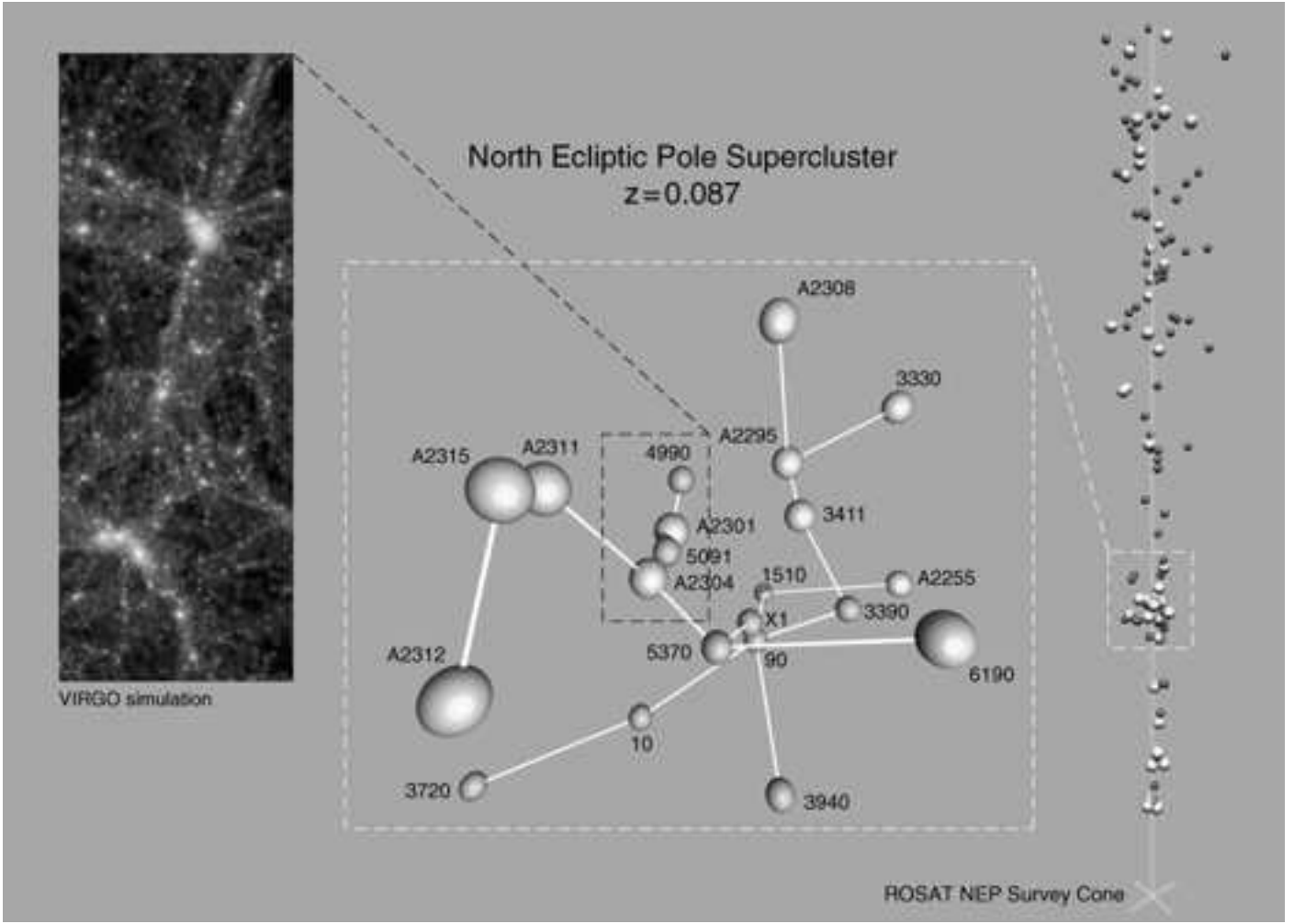}
\figcaption{RIGHT: Three-dimensional view of the ROSAT NEP Survey
volume out to $z=0.3$ with a dashed box outlining the NEPSC.  Yellow
spheres represent clusters and groups of galaxies, while red spheres
represent AGN.  The Earth is at the origin with the long, redshift
axis pointed towards the NEP.  An animated ``fly-through'' of the NEP
Survey volume is available at \em http://www.ifa.hawaii.edu/\~{
}mullis/nep3d.html\em.  MIDDLE: Close-up view of the NEPSC where
galaxy clusters (yellow spheres) and galaxy groups (smaller green
spheres) are linked with a minimum spanning tree of segments.  Objects
from the ROSAT NEP survey are identified with strictly numerical
labels, Abell clusters are prefixed with ``A'', and ``X1'' marks the
faint group from Burg et al.\ 1992.  The scale varies in this
three-dimensional perspective.  LEFT: A filamentary complex of
clusters and groups from a Virgo consortium $\Lambda$+CDM ($z=0$)
numerical simulation (Jenkins et al.\ 1998) that resembles the $20.7
~h_{50}^{-1}\ $Mpc long, A2304-NEP4990 filament.\break
\mbox{} \vskip .3in
\bf A high-resolution color version of this figure is available 
at \bf \hspace*{\fill} \break
\bf http://www.ifa.hawaii.edu/\~{}mullis/papers/nepsc.ps.gz \bf}

\end{figure*}

\end{document}